\documentclass{PoS}

\title{Transverse momentum and multiplicity fluctuations in the Be+Be energy scan from NA61/SHINE}

\ShortTitle{$p_{T}$ and N fluctuations in Be+Be energy scan from NA61/SHINE}

\author{\speaker{Tobiasz Czopowicz} for the NA61/SHINE Collaboration\\
        Warsaw University of Technology\\
        E-mail: \email{tobiasz.czopowicz@cern.ch}}

\abstract{
    The NA61/SHINE experiment aims to discover the Critical Point of strongly interacting matter
    and study the properties of the onset of deconfinement. These goals are to be achieved by
    performing a two-dimensional phase diagram (T-$\mu_{B}$) scan by measurements of hadron
    production properties in proton-proton, proton-nucleus and nucleus-nucleus interactions as
    a function of collision energy and system size. Close to the Critical Point an increase of
    fluctuations is predicted.

    This contribution presents preliminary results on transverse momentum and multiplicity fluctuations
    expressed in terms of strongly intensive quantities from the Be+Be
    energy scan. The data are fully corrected for contributions from non-target interactions.
    The Be+Be results are compared with NA61/SHINE measurements from the p+p energy scan,
    with NA49 results from central Pb+Pb collisions as well as with model predictions.
}

\FullConference{
    9th International Workshop on Critical Point and Onset of Deconfinement - CPOD2014,\\
    17-21 November 2014\\
    ZiF (Center of Interdisciplinary Research), University of Bielefeld, Germany
}

\begin{document}

\section{Introduction and motivation}
    NA61/SHINE \cite{NA61} is a fixed-target experiment located at the Super Proton Synchrotron (SPS)
    in the North Area at CERN. It measures hadron production in proton-proton, proton-nucleus
    and nucleus-nucleus collisions at various beam momenta (13$A$ -- 158$A$ GeV/c).

    The main goal of the experiment is to discover the Critical Point of strongly interacting
    matter and study the properties of the onset of deconfinement.

    One of the tools for studying the phase transition region and search for the Critical Point (CP)
    is the analysis of event-by-event fluctuations of multiplicity and kinematic characteristics.
    These fluctuations and correlations may serve as a signature of the onset of deconfinement
    (close to the phase transition the Equation of State changes rapidly which can impact the energy
    dependence of fluctuations) and can help to locate the CP of strongly interacting matter.

    The NA49 collaboration \cite{NA49} has studied proton-proton and nucleus-nucleus collisions.
    In their system-size scan with p, C, Si and Pb beams at 158$A$ GeV/c \cite{NA49_phi_system_size} an
    increase of fluctuations for medium-size systems, consistent with CP predictions \cite{SRS} has
    been observed.

\begin{figure}[ht]
        \centering
        \includegraphics[width=.35\textwidth]{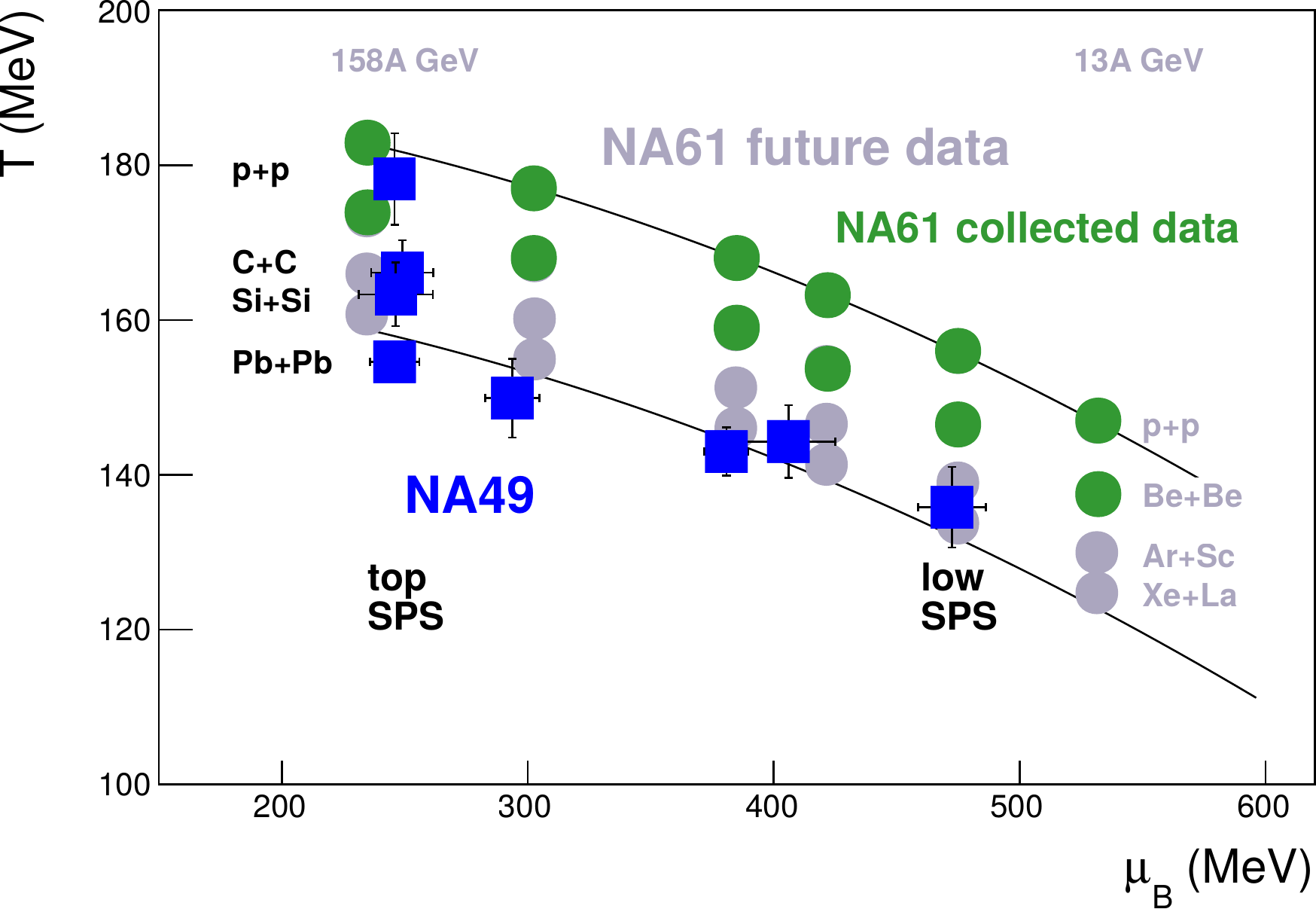} \qquad
        \caption{
            NA61/SHINE data taking plan. Each data set is marked as a point on the $T-\mu_{B}$
            phase diagram of strongly interacting matter. Fitted (NA49 - blue squares) and
            predicted (NA61/SHINE - gray/green circles) chemical freeze-out points based on \cite{becattini}.
        }
        \label{fig:phase-diagram}
    \end{figure}

    NA61/SHINE is performing a systematic two-dimensional scan of the phase diagram (Fig.~\ref{fig:phase-diagram})
    in which the CP (if it exists) is expected to show up as a hill-like structure in the
    energy-system size dependence of fluctuations.

\section{The analysis}
    Four data sets of $\mathrm{{}^{7}Be+{}^{9}Be}$ collisions recorded by NA61/SHINE in 2011 and the winter 2012/13
    have been analyzed. Statistics are shown in Table~\ref{tbl:datasets}.
    \begin{table}[ht]
        \caption{
            Number of analyzed Be+Be interactions before and after quality cuts.
        }
        \label{tbl:datasets}
        \centering
        \begin{tabular}{c|c|c c|c c}
            \hline \hline
            $p_{beam}$ & $\sqrt{s_{NN}}$ & \multicolumn{2}{|c|}{target inserted} & \multicolumn{2}{|c}{target removed} \cr
            [GeV/c]    & [GeV]           & all           & selected              & all           & selected            \cr \hline
            20         & 6.27            & 3.2 M         & 1.2 M                 & 305 k         & 2.5 k               \cr
            40         & 8.73            & 3 M           & 0.9 M                 & 319 k         & 1.8 k               \cr
            75         & 11.94           & 4.2 M         & 1.2 M                 & 433 k         & 2.4 k               \cr
            150        & 16.83           & 2.5 M         & 0.9 M                 & 283 k         & 2.4 k               \cr \hline \hline
        \end{tabular}
    \end{table}

    A number of selection criteria were used to ensure good quality inelastic
    Be+Be collisions produced by a well-defined beam particle without other beam particles close in time.
    Reconstructed particles must originate from a well reconstructed main vertex and leave sufficient
    numbers of points inside the TPCs. Electrons and positrons were excluded.
    The analysis was limited to particles with transverse momentum less than 1.5 GeV/c.

    \begin{figure}[ht]
        \centering
        \includegraphics[width=.3\textwidth]{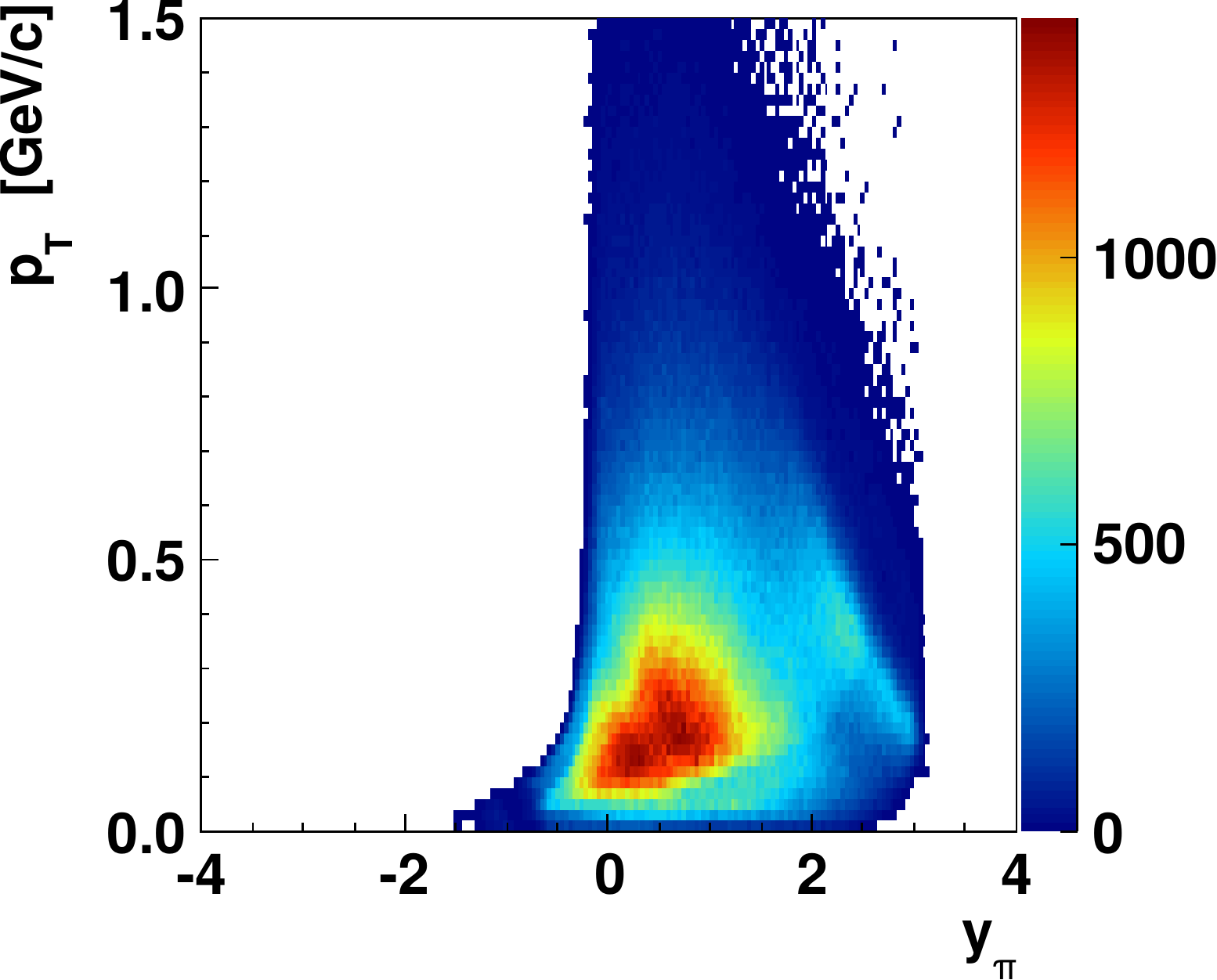} \qquad
        \includegraphics[width=.3\textwidth]{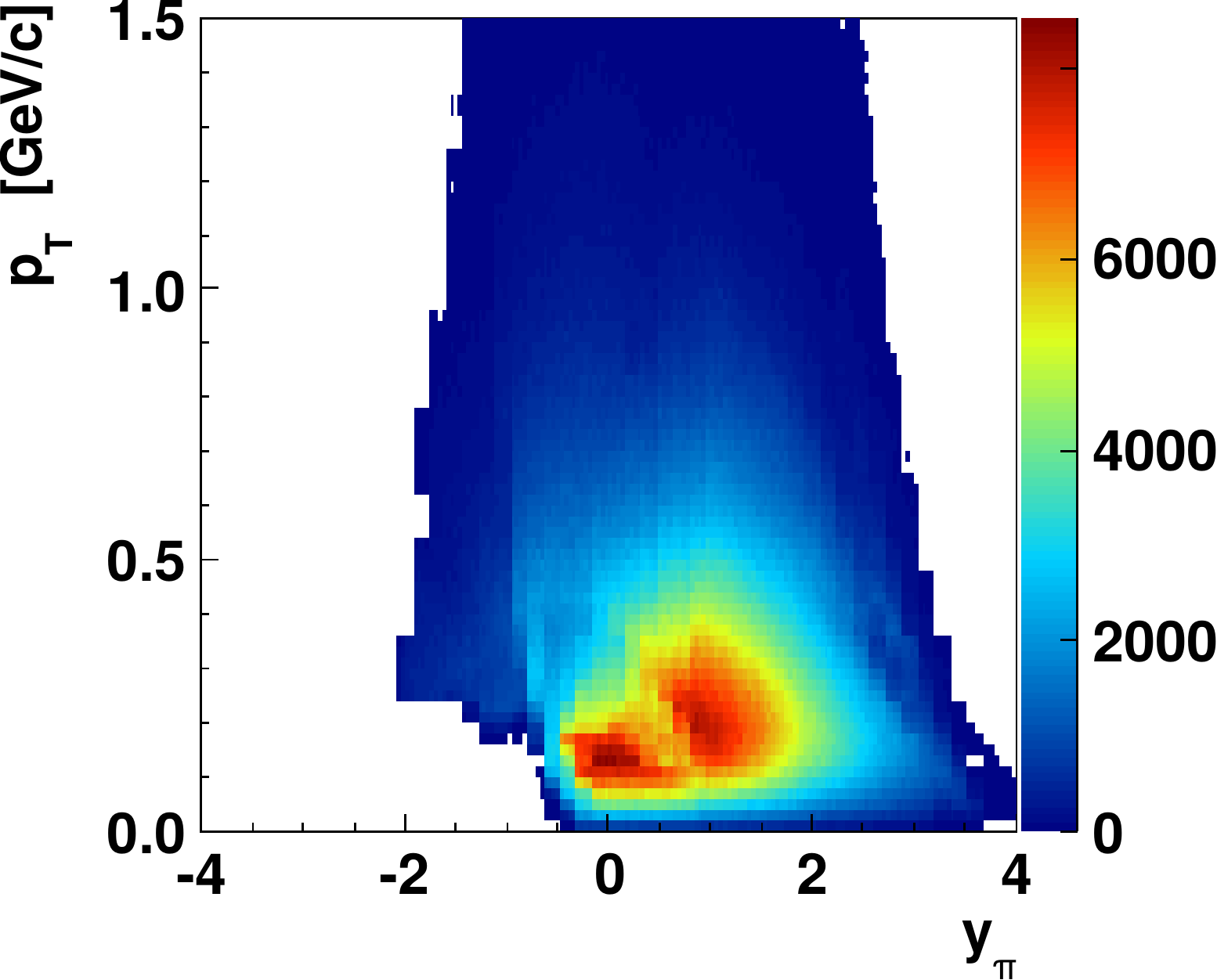} \\[1em]
        \includegraphics[width=.3\textwidth]{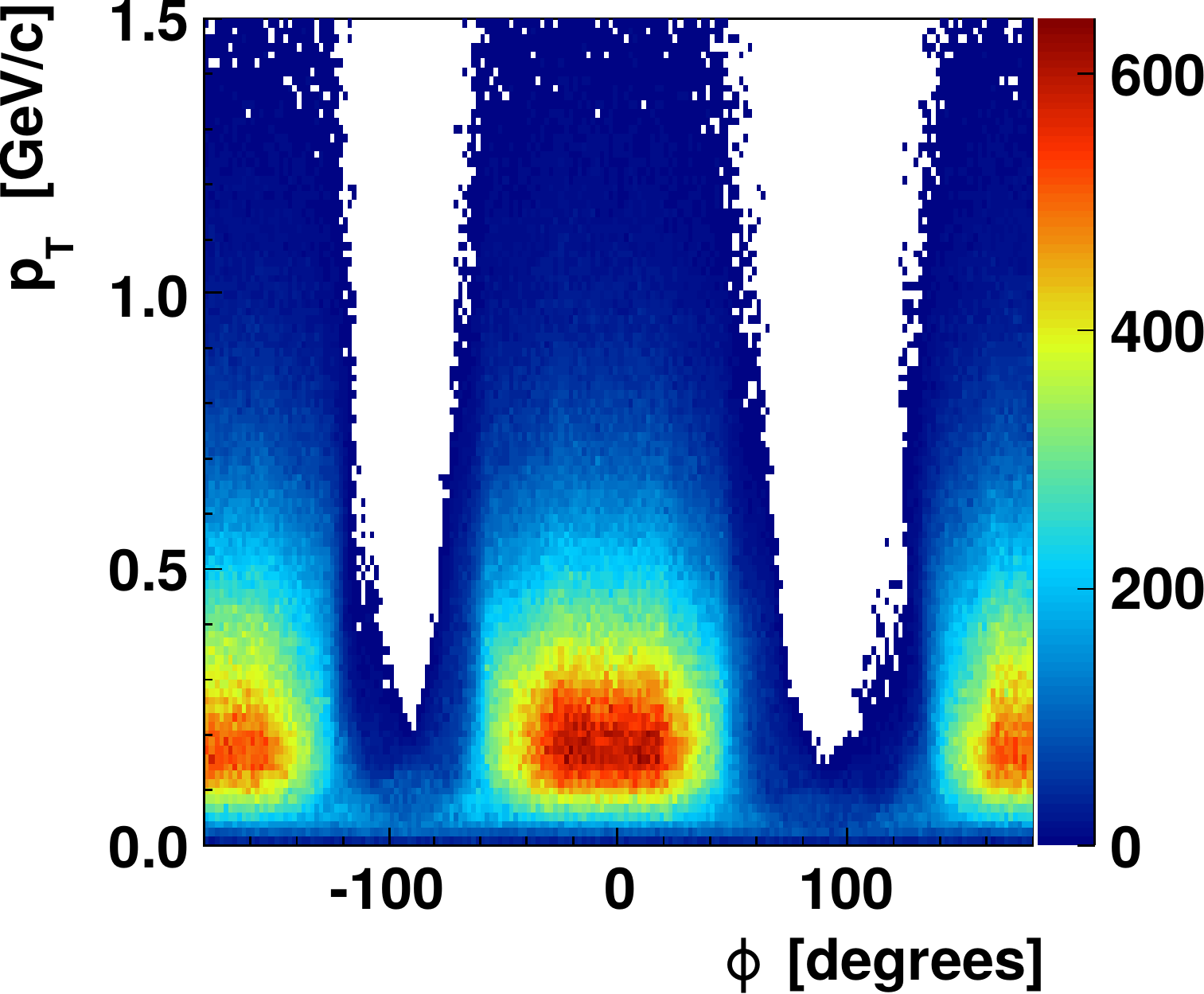} \qquad
        \includegraphics[width=.3\textwidth]{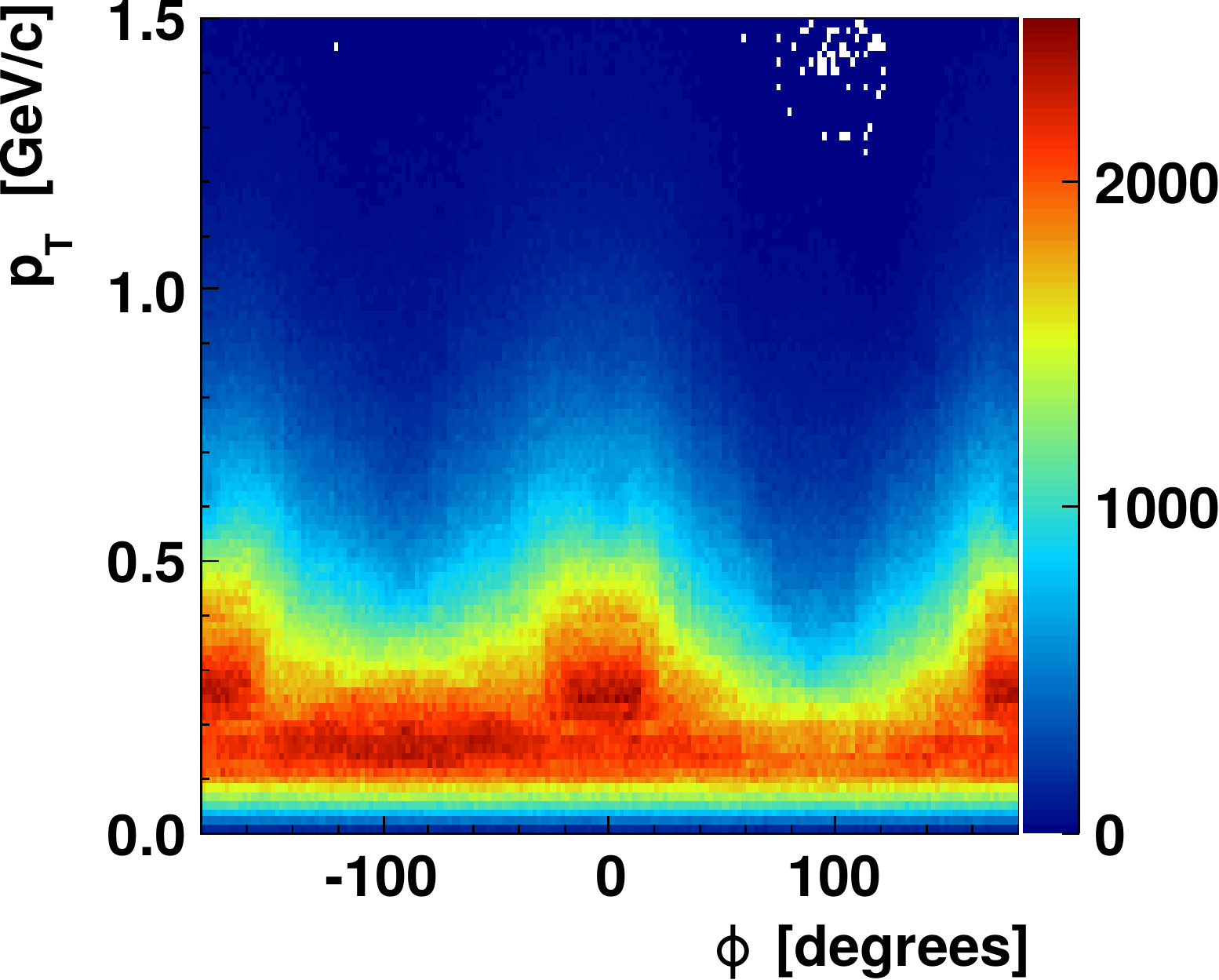}
        \caption{
            Distribution of accepted particles in center-of-mass rapidity (top) and azimuthal angle
            (bottom) as a function of transverse momentum for Be+Be collisions
            at minimum (20$A$ GeV/c, left) and maximum (150$A$ GeV/c, right) beam momentum.
        }
        \label{fig:acceptance}
    \end{figure}

    Centrality was determined from the energy deposited in the NA61/SHINE forward calorimeter --
    the Particle Spectator Detector (PSD) (see E.Kaptur, this proceedings).

    The fluctuation results are corrected for contributions from non-target interactions. For this purpose NA61/SHINE
    acquired data with both target inserted and removed. Then the contribution from non-target interactions is
    subtracted in the analysis.

    Corrections for detector effects and trigger bias are estimated to be small but still under
    investigation.

\section{Fluctuation Measures}
    As fluctuations in high-energy heavy-ion collisions may have different sources, it is important
    to select the fluctuations of interest. Despite a good PSD performance, it is
    impossible to determine exactly volume of the created matter on an event-by-event basis. However, it is
    possible to construct fluctuation measures which do not depend on volume and its fluctuations
    (strongly intensive measures) in statistical models in the grand canonical ensamble (GCE).

    \subsection{Scaled variance}
        The most natural way to describe fluctuations of the quantity $A$ is to measure the variance of its
        distribution. This is an extensive quantity, i.e. proportional to the system size.
        However, the ratio of two extensive variables is independent of system size (intensive measure).
        So the scaled variance
        \begin{equation}
            \omega[A] = \frac{Var(A)}{\langle A \rangle}
        \end{equation}
        does not depend on the average volume, but it depends on volume fluctuations.

        The scaled variance is a measure of multiplicity fluctuations used extensively in the past and
        also studied by NA61/SHINE.

    \subsection{Strongly intensive measures}
        It is possible to define a combination of scaled variances of two quantities, $A$ and $B$,
        in such a way, that their dependence on average volume and volume fluctuations cancels out.
        Such a measure is called {\it strongly intensive}.

        There are two families of strongly intensive fluctuation measures \cite{strongly}:
        \begin{eqnarray}
            &\Delta[A,B] = \frac{1}{C_{\Delta}} \biggl[ \langle B \rangle \omega[A] -
                        \langle A \rangle \omega[B] \biggr] \\
            &\Sigma[A,B] = \frac{1}{C_{\Sigma}} \biggl[ \langle B \rangle \omega[A] +
                        \langle A \rangle \omega[B] - 2 \bigl( \langle AB \rangle -
                        \langle A \rangle \langle B \rangle \bigr) \biggr]
        \end{eqnarray}
        In case of transverse momentum fluctuations:
        $$
            A = P_{T} = \sum\limits_{i=1}^{N} p_{T_{i}}, \qquad B = N, \qquad
            C_{\Delta} = C_{\Sigma} = \langle N \rangle \omega[p_{T}].
        $$

        They are both equal to 0 if there are no fluctuations and they are both
        equal to 1 for the Independent Particle Model \cite{new_norm}. However, there
        is one important difference between the two measures -- $\Sigma[P_{T},N]$ contains a
        correlation term, $\langle AB \rangle$.

        To compare with published NA49 data, the strongly intensive $\Phi_{p_T}$ measure \cite{Phi_measure} is used:
        \begin{equation}
            \Phi_{p_T} = \sqrt{\overline{p_T} \omega[p_T]} \left[\sqrt{\Sigma[P_{T},N]}-1\right],
        \end{equation}

\section{Results}
\begin{figure}[ht]
        \centering
        \includegraphics[width=.3\textwidth]{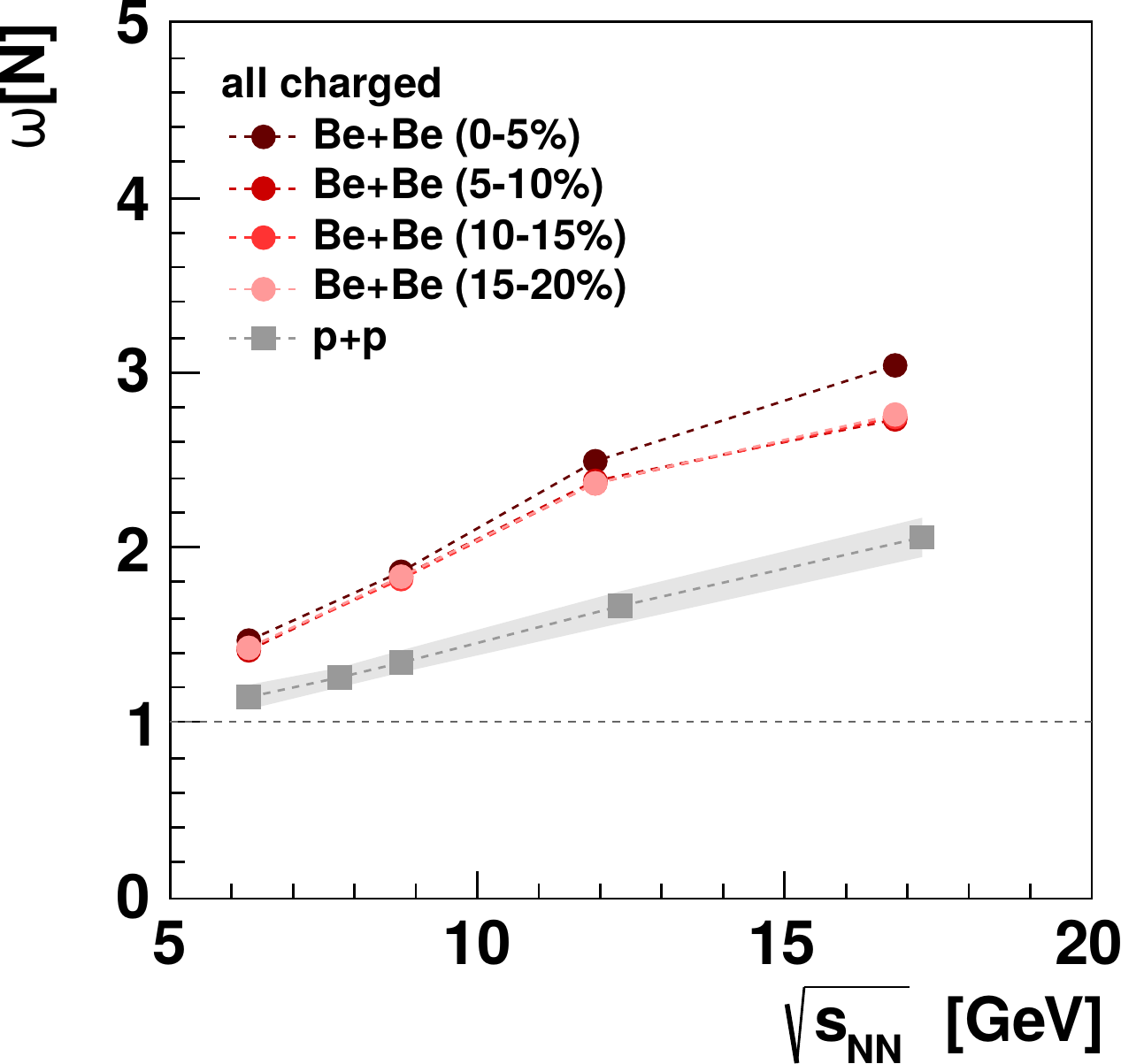} \quad
        \includegraphics[width=.3\textwidth]{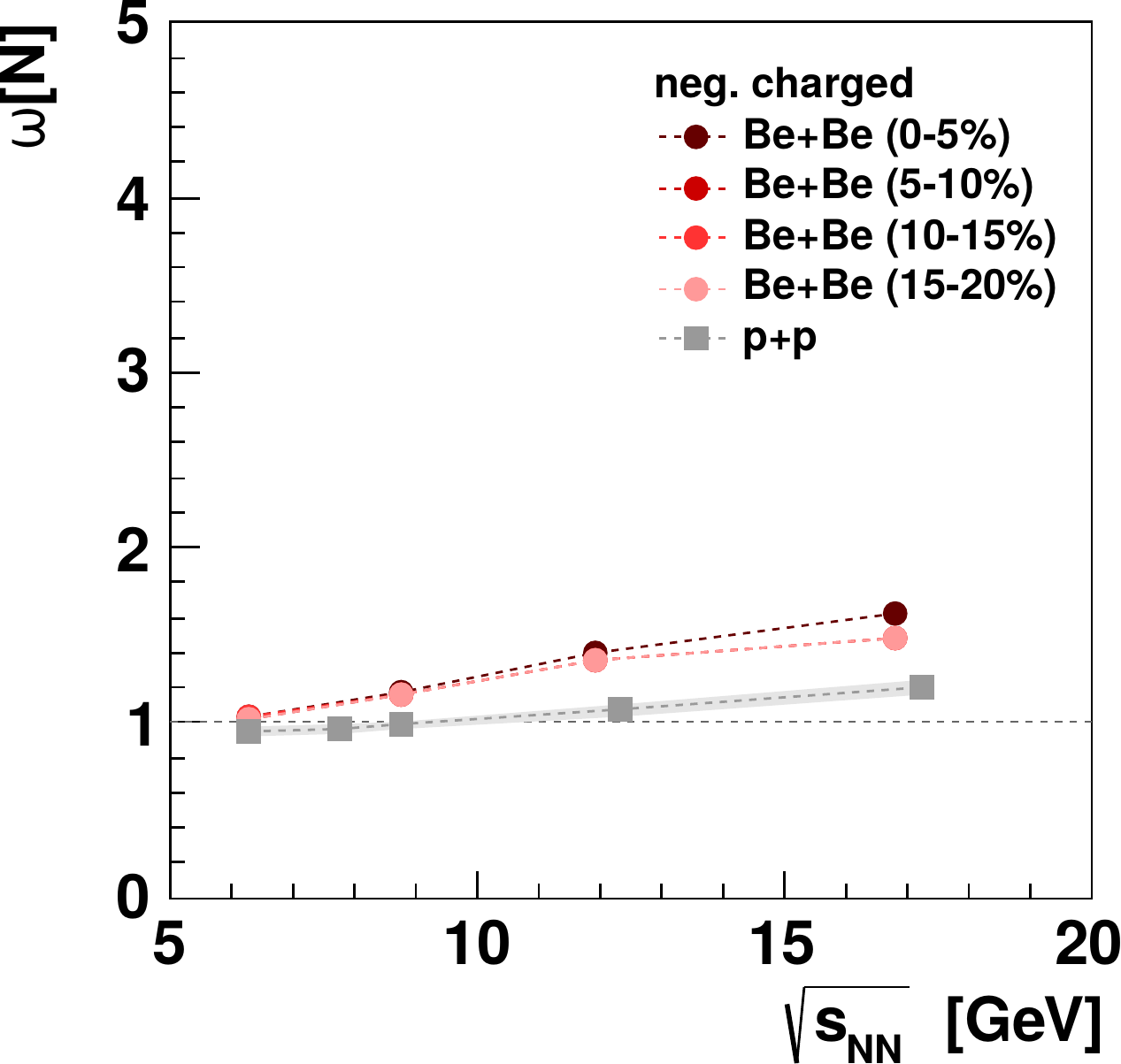} \quad
        \includegraphics[width=.3\textwidth]{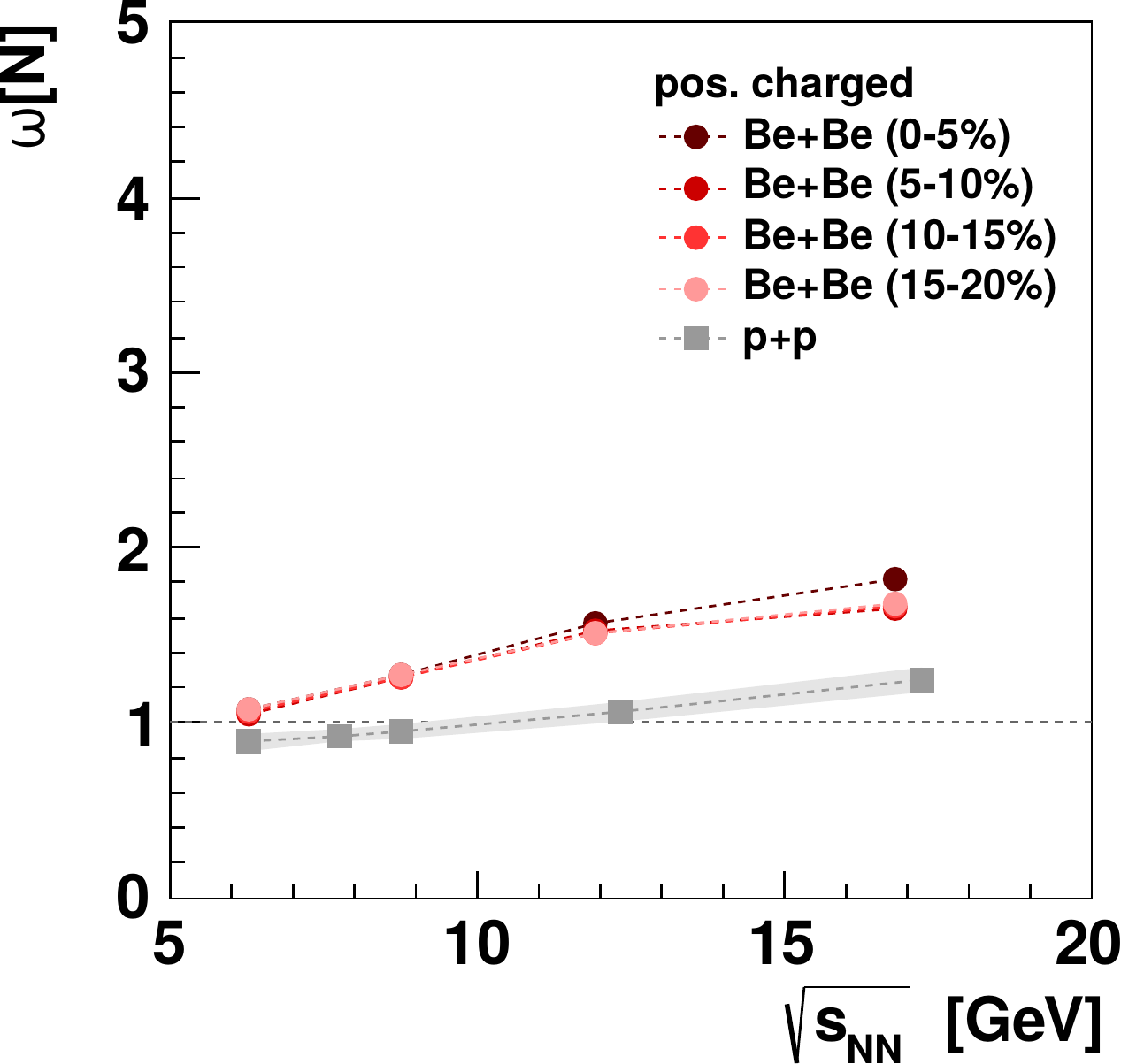}
        \caption{
            Scaled variance of the multiplicity distribution for Be+Be collisions
            of several collision centralities and energies
            compared with the correcponding results from the energy scan with inelastic p+p interactions.
        }
        \label{fig:omega}
    \end{figure}

    Figure~\ref{fig:omega} presents results for the scaled variance of the multiplicity distribution
    from Be+Be collisions for four energies, compared with measurements of NA61/SHINE from the energy scan
    of inelastic p+p interactions. $\omega$ increases with energy -- probably due to KNO scaling
    (variance of the multiplicity distribution increases faster than its mean).
    Larger values for Be+Be than for p+p collisions are
    probably due to volume fluctuations because $\omega$ is not a strongly intensive measure.

\begin{figure}[ht]
        \centering
        \includegraphics[width=.3\textwidth]{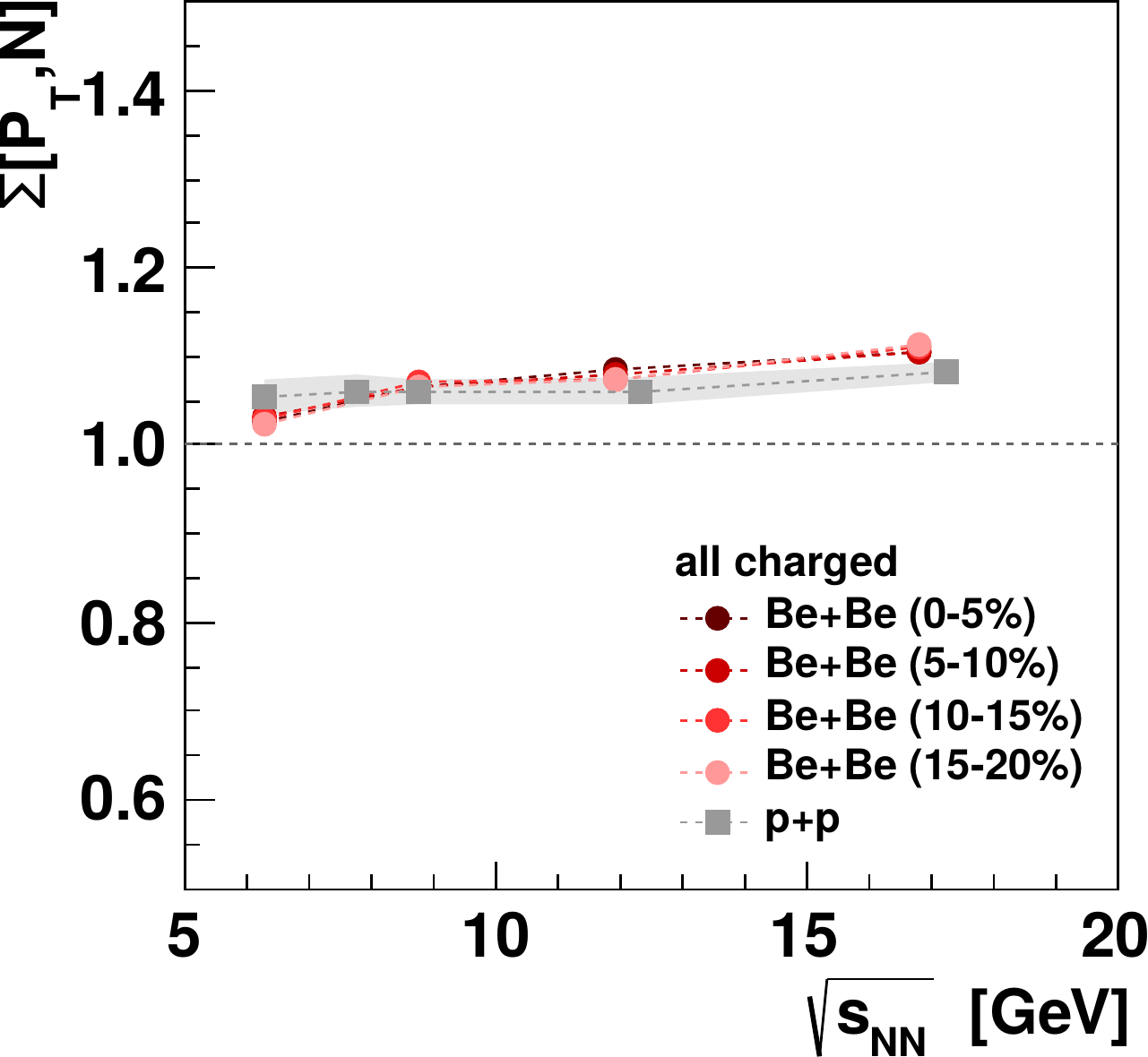} \quad
        \includegraphics[width=.3\textwidth]{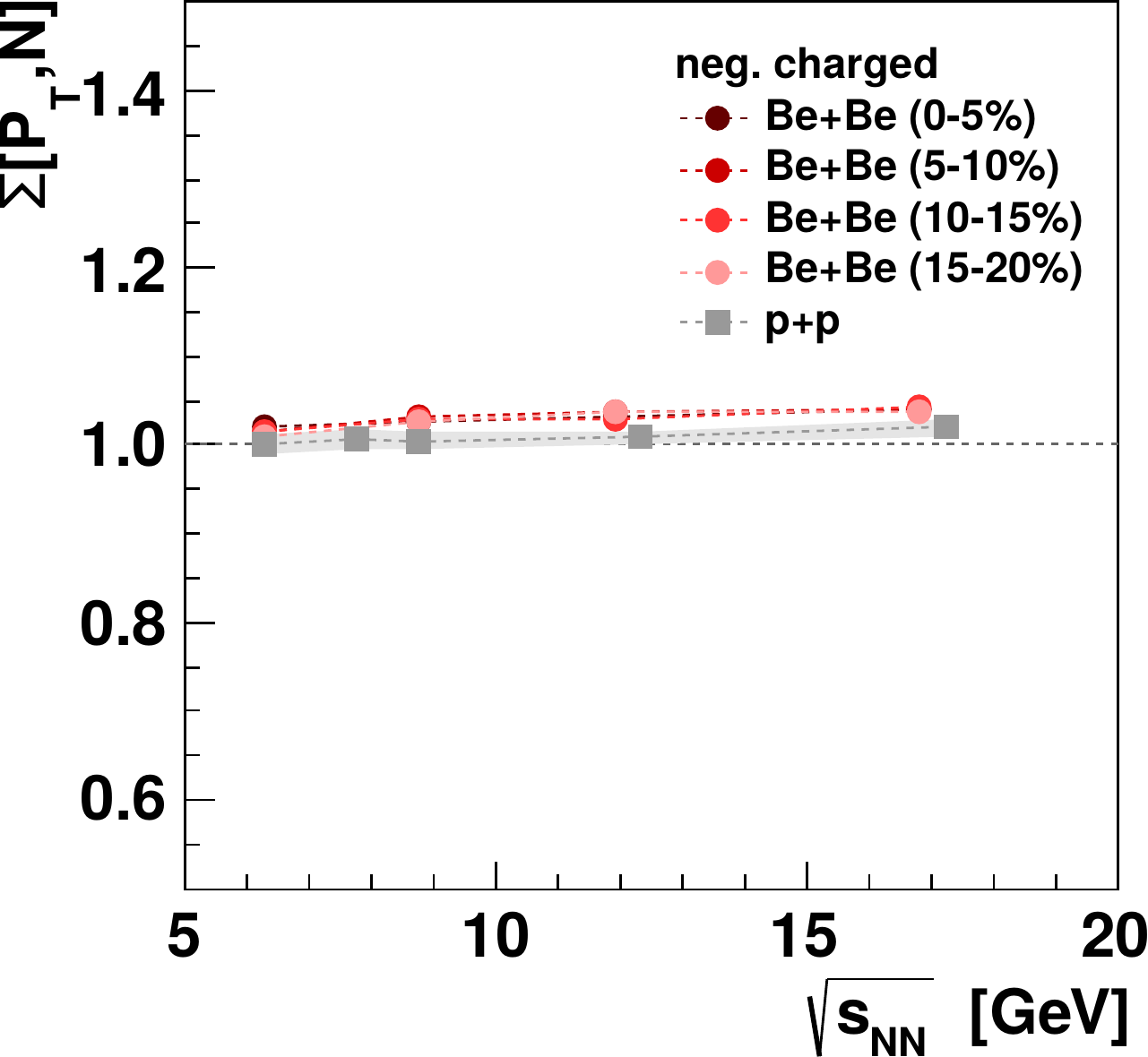} \quad
        \includegraphics[width=.3\textwidth]{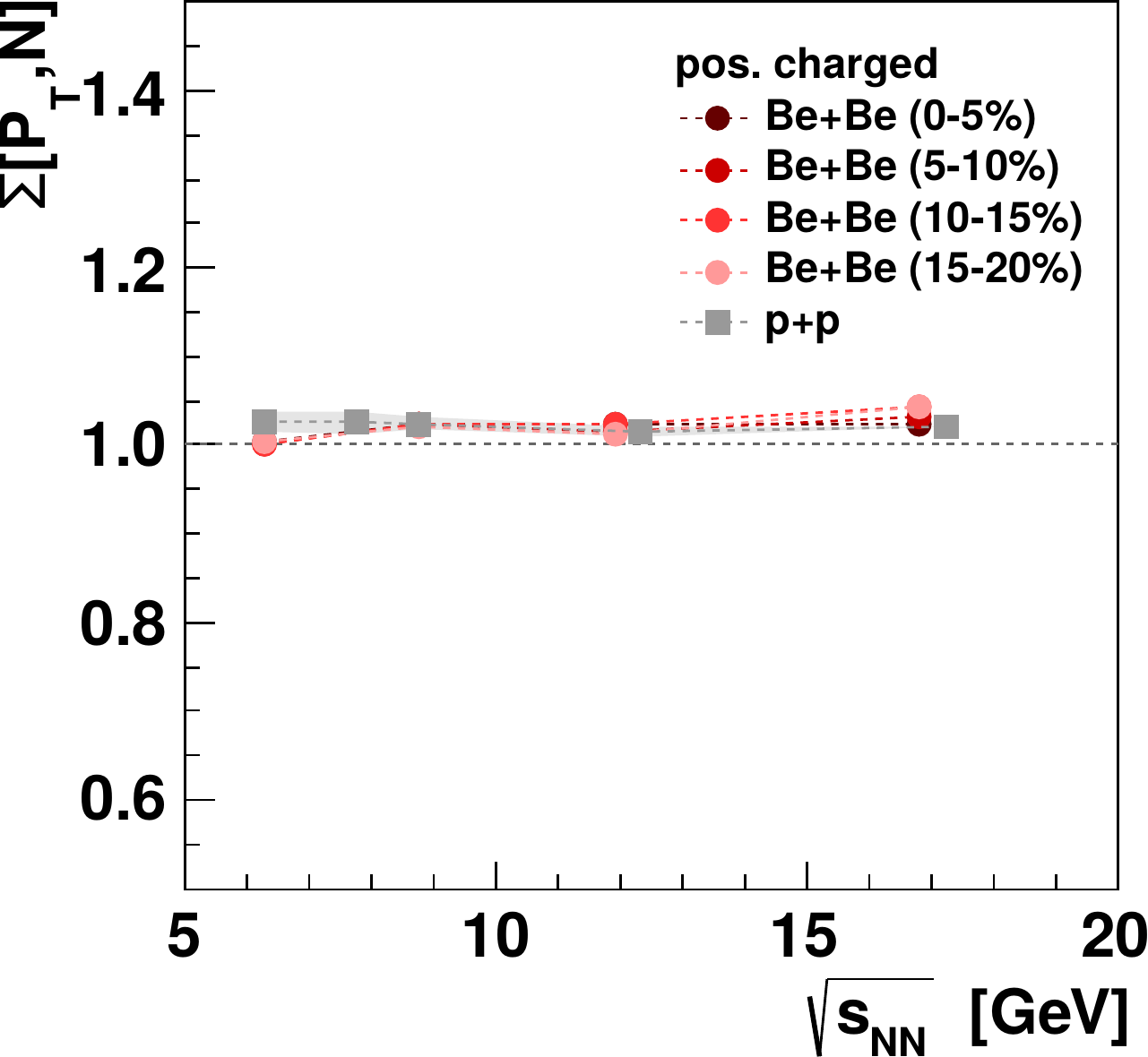} \\[1em]
        \includegraphics[width=.3\textwidth]{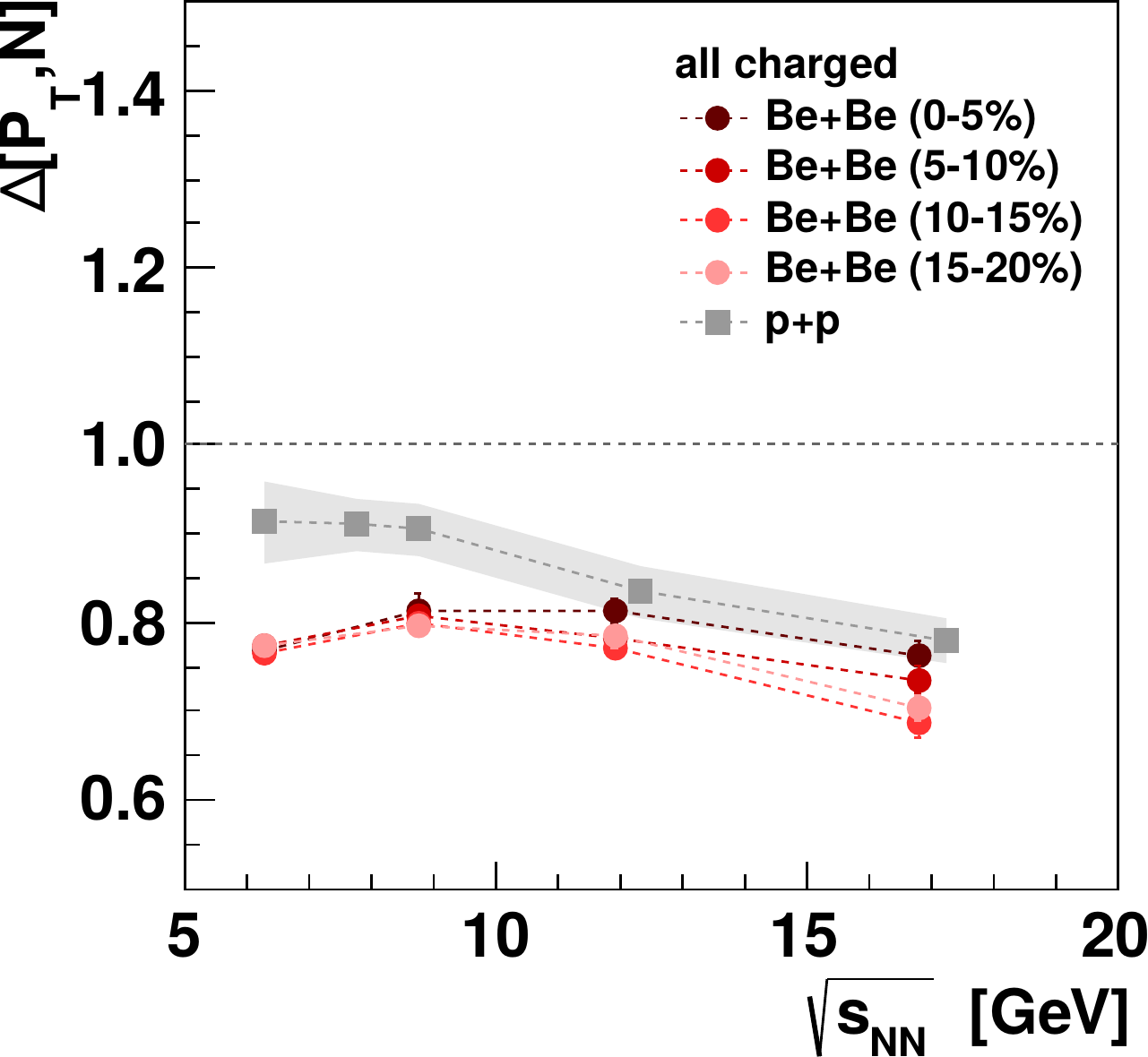} \quad
        \includegraphics[width=.3\textwidth]{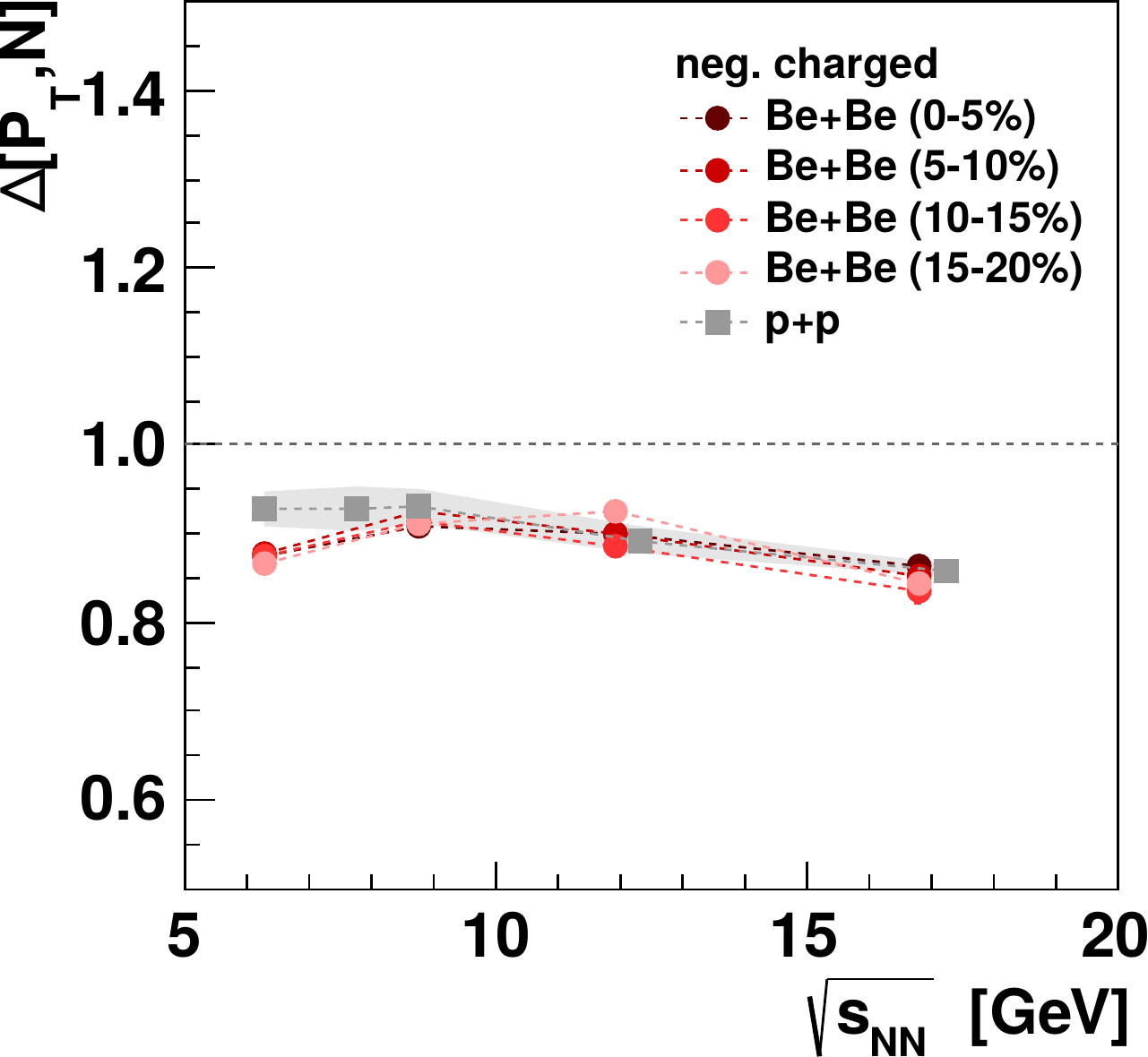} \quad
        \includegraphics[width=.3\textwidth]{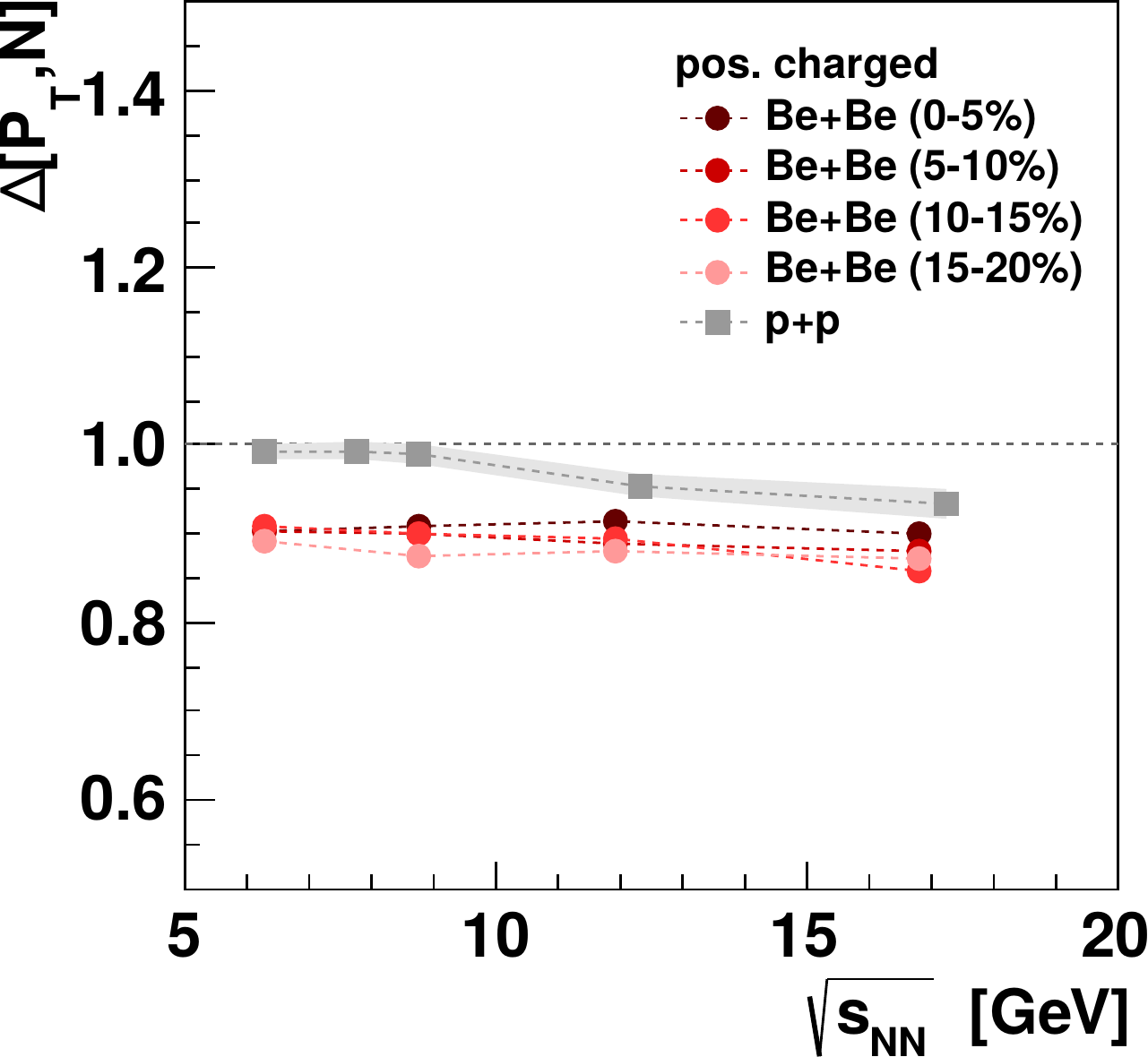} \\[1em]
        \includegraphics[width=.3\textwidth]{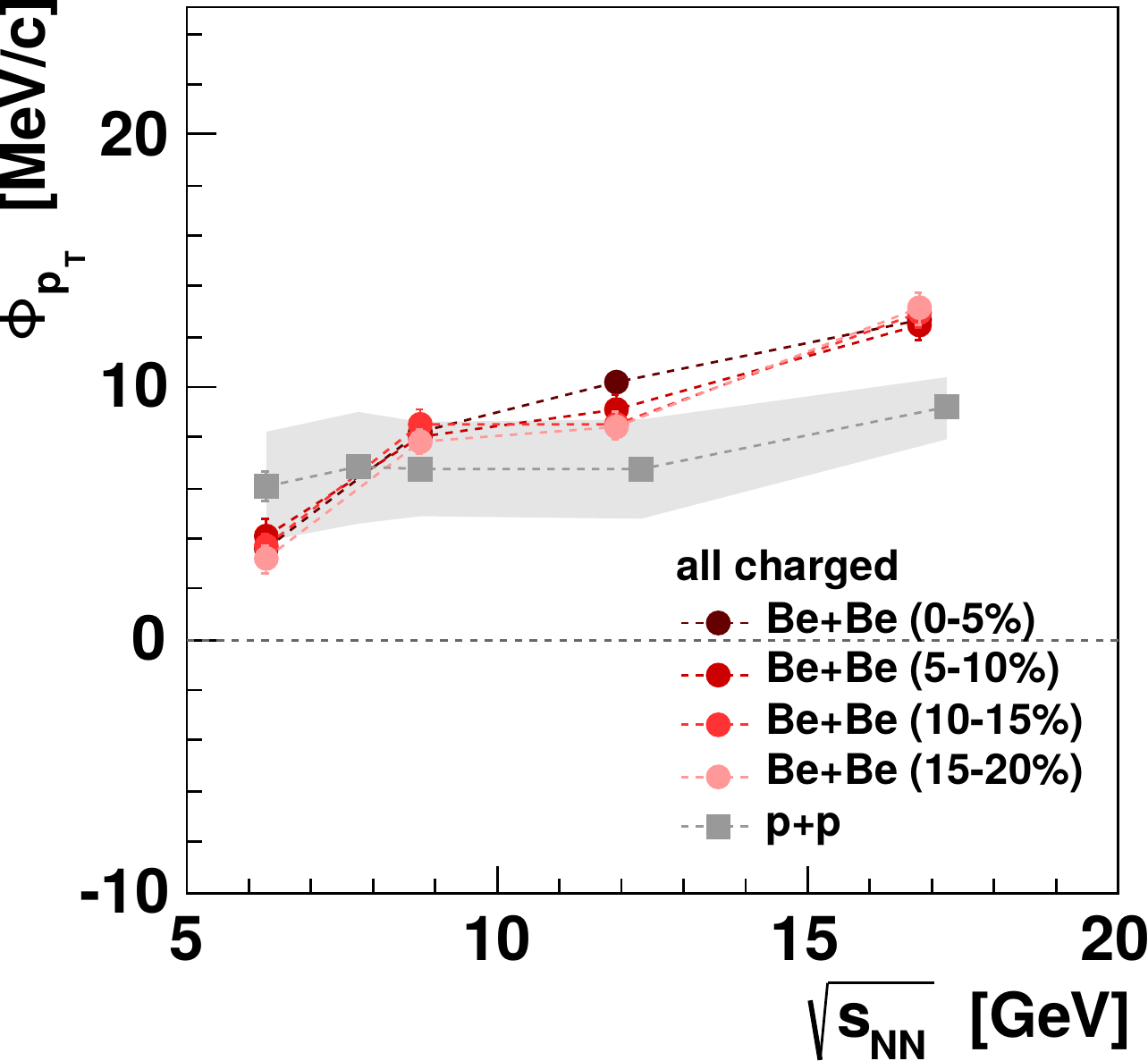} \quad
        \includegraphics[width=.3\textwidth]{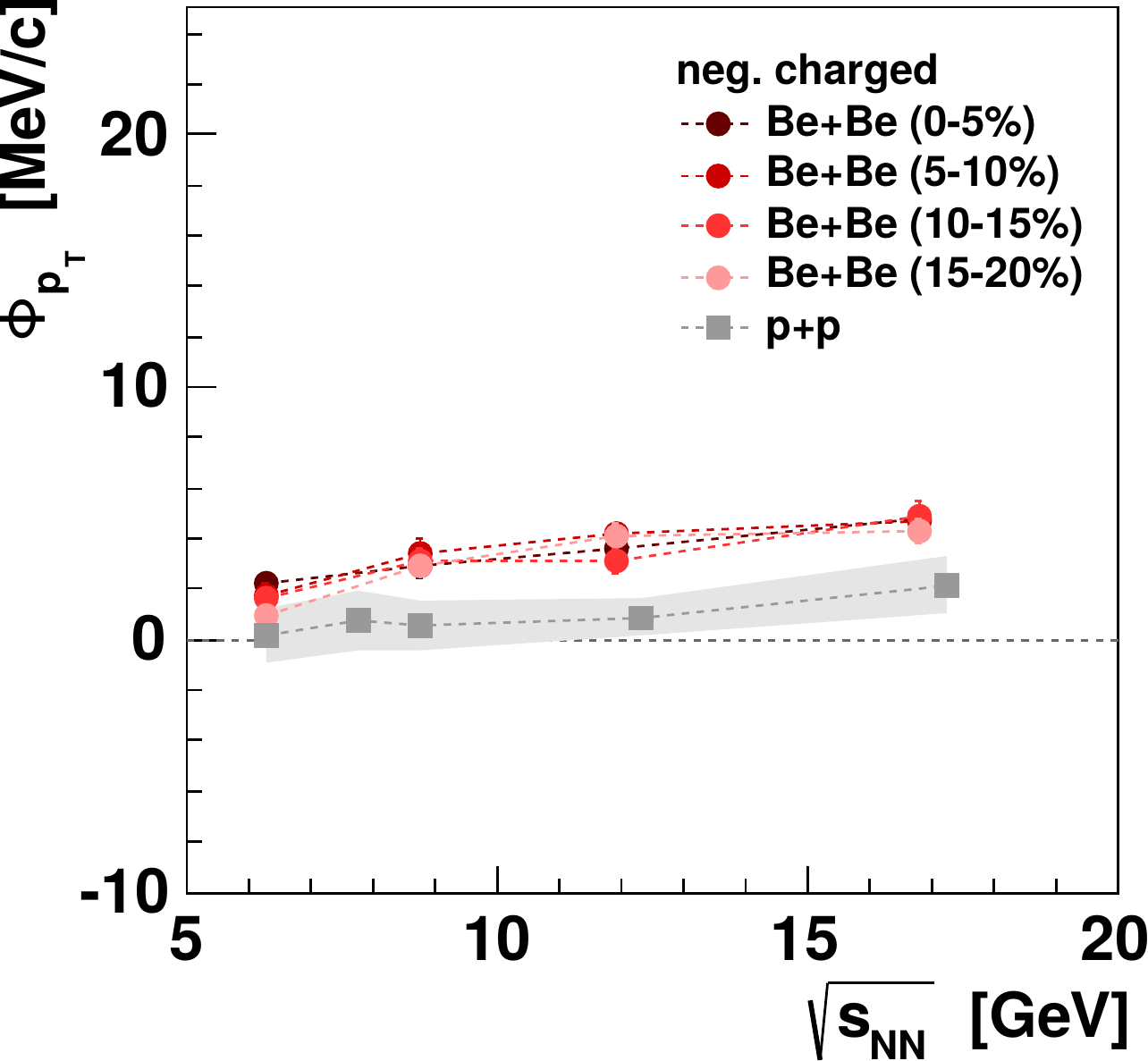} \quad
        \includegraphics[width=.3\textwidth]{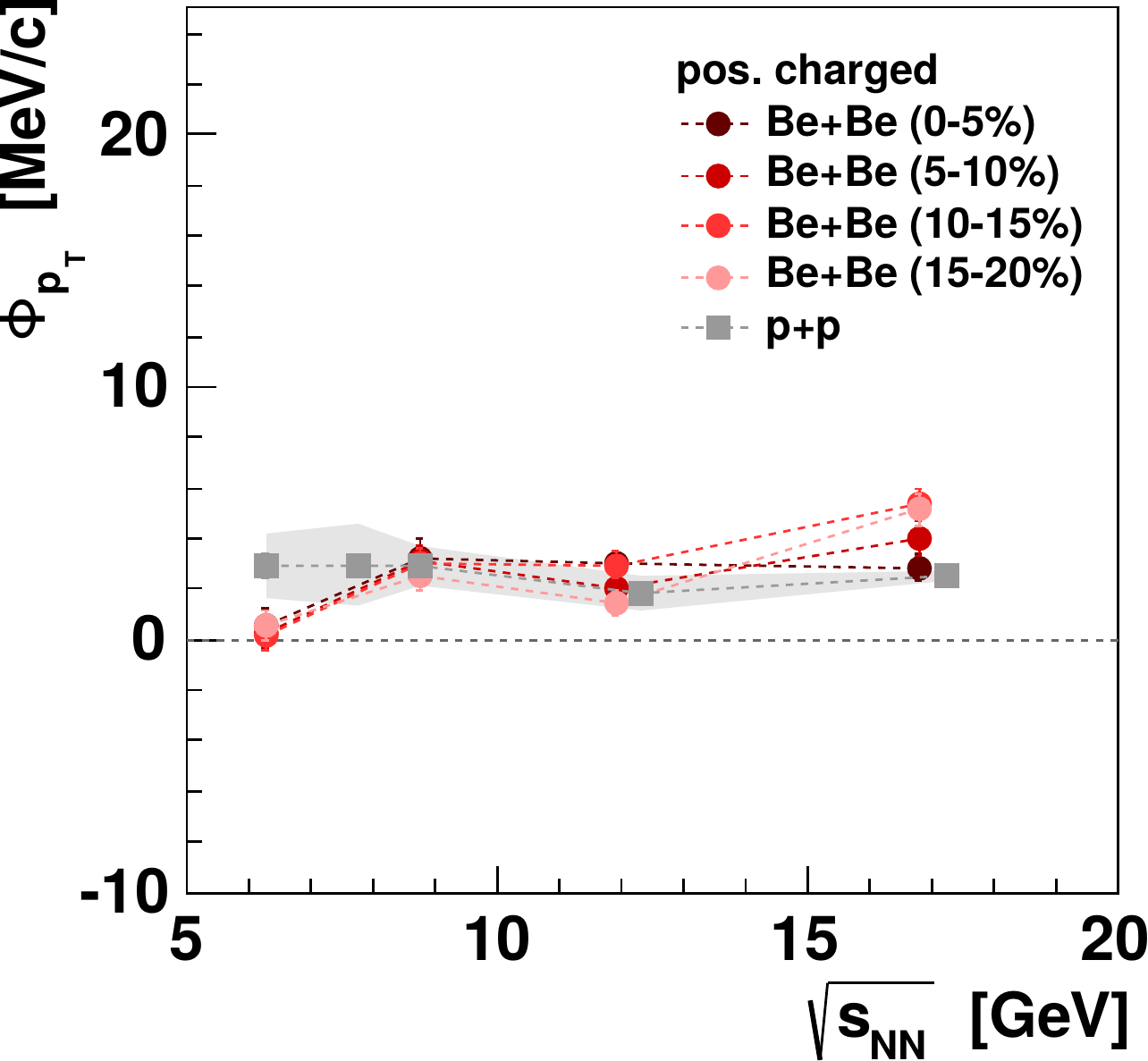}
        \caption{
            $\Delta[P_{T},N]$, $\Sigma[P_{T},N]$ and $\Phi_{p_T}$ for Be+Be collisions
            of several collision centralities and energies
            compared with the correcponding results from the energy scan with inelastic p+p interactions.
        }
        \label{fig:strongly}
    \end{figure}

    Figure~\ref{fig:strongly} shows results for transverse momentum fluctuations measured by the
    strongly intensive measures $\Delta[P_{T},N]$, $\Sigma[P_{T},N]$ and $\Phi_{p_T}$ for
    Be+Be interactions of several collision centralities and energies. They are compared with
    measurements of NA61/SHINE from the energy scan of inelastic p+p interactions.

    Transverse momentum fluctuations in Be+Be and p+p interactions show no structures which
    could be related to the CP. Be+Be results are close to those from inelastic p+p reactions.
    The weak dependence on centrality in Be+Be shows that the fluctuations do not depend on the collision volume.
    The difference between $\Delta[P_{T},N]$ and $\Sigma[P_{T},N]$ might be due to Bose-Einstein
    and/or $P_{T}-N$ correlations \cite{KG_MG}.

\subsection{Comparison with NA49}
    As it is impossible to correct fluctuations for the acceptance, additional cuts were applied to compare
    NA61/SHINE results on $\Phi_{p_T}$ with the published NA49 measurements at 158$A$ GeV/c beam energy.
    The NA49 analysis was limited to the forward rapidity region ($1.1 < y_{\pi} < 2.6$) since
    the high density of tracks did not allow reliable measurements in the mid-rapidity region.
    Moreover, a common azimuthal angle acceptance was used for all systems.

    \begin{figure}[ht]
        \centering
        \includegraphics[width=.45\textwidth]{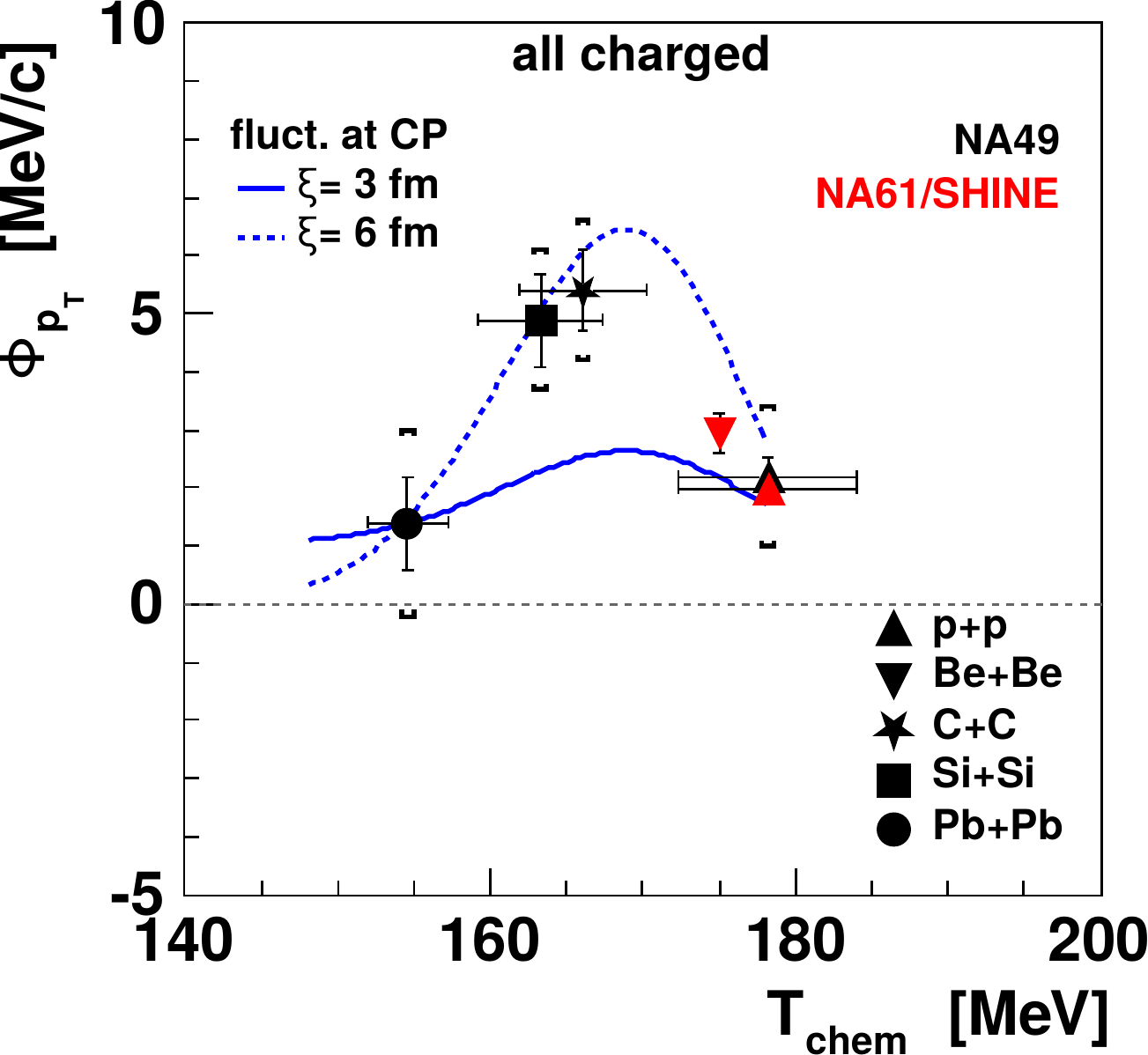} \quad
        \caption{
            System size ($T_{chem}$) dependence of transverse momentum fluctuations at 158$A$ GeV/c
            (150$A$ GeV/c for Be+Be). NA61/SHINE results (p+p, central Be+Be) are compared to NA49 data
            (p+p, semi-central C+C, semi-central Si+Si, central Pb+Pb) in the same NA49 acceptance
            (see \cite{NA49_phi_system_size}).
        }
        \label{fig:comparison}
    \end{figure}

    The comparison is shown in Fig.~\ref{fig:comparison}. Results of NA61/SHINE from p+p interactions agree
    with corresponding measurements from NA49. Also the result for Be+Be collisions is consistent with the
    system size dependence suggested by NA49.

    The ongoing Ar+Sc energy scan is expected to give more insight into the existence and location of the CP.

\section*{Acknowledgements}
    This work was supported by the polish National Science Centre (grant 2012/04/M/ST2/00816),
    the Hungarian Scientific Research Fund (grants OTKA 68506 and 71989),
    the Polish Ministry of Science and Higher Education (grants 667/N-CERN/2010/0, NN 202 48 4339
    and NN 202 23 1837),
    the Federal Agency of Education of the Ministry of Education and Science of the Russian Federation
    (grant RNP 2.2.2.2.1547),
    the Russian Academy of Science and the Russian Foundation for Basic Research
    (grants 08-02-00018,09-02-00664 and 12-02-91503-CERN),
    the Ministry of Education, Culture, Sports, Science and Technology,
    Japan, Grant-in-Aid for Scientific Research (grants 18071005, 19034011, 19740162, 20740160 and 20039012),
    the German Research Foundation (grant GA 1480/2-1),
    Bulgarian National Scientific Fondation (grant DDVU 02/19/ 2010),
    Ministry of Education and Science of the Republic of Serbia (grant OI171002),
    Swiss Nationalfonds Foundation (grant 200020-117913/1)
    and ETH Research Grant TH-01 07-3.


\begin{thebibliography}{99}
	\bibitem{NA61}
		N.~Abgrall {\it et al.}  [NA61 Collaboration],
		%``NA61/SHINE facility at the CERN SPS: beams and detector system'',
		JINST {\bf 9}, P06005 (2014)

	\bibitem{NA49}
		S.V.~Afanasiev {\it et al.} [NA49 Collaboration],
		%``The NA49 large acceptance hadron detector'',
		Nucl.\ Instrum.\ Meth.\ A {\bf 430}, 210 (1999).

	\bibitem{NA49_phi_system_size}
		T. Anticic {\it et al.} [NA49 Collaboration],
		%``Transverse momentum fluctuations in nuclear collisions at 158-A-GeV'',
		Phys.\ Rev.\ C {\bf 70} (2004) 034902.
		%[arXiv:hep-ex/0311009]

	\bibitem{SRS}
		M.~A.~Stephanov, K.~Rajagopal, E.~V.~Shuryak,
		%``Event-by-event fluctuations in heavy ion collisions and the QCD Critical Point'',
		Phys.\ Rev.\ D {\bf 60} (1999) 114028.
		%[arXiv:hep-ph/9903292]

	\bibitem{becattini}
		F.~Becattini, J.~Manninen, M.~Gazdzicki,
		%``Energy and system size dependence of chemical freeze-out in relativistic nuclear collisions'',
		Phys.\ Rev.\ C {\bf 73} (2006) 044905.
		%[arXiv:hep-ph/0511092]

	\bibitem{strongly}
		M.~I.~Gorenstein, M.~Gazdzicki,
		%``Strongly Intensive Quantities'',
		Phys.\ Rev.\ C {\bf 84} (2011) 014904.
		%[arXiv:nucl-th/1101.4865]

	\bibitem{new_norm}
		M.~Gazdzicki, M.~I.~Gorenstein and M.~Mackowiak-Pawlowska,
		%``Normalization of strongly intensive quantities'',
		Phys.\ Rev.\ C {\bf 88} (2013) 2,  024907
		%[arXiv:1303.0871 [nucl-th]].

	\bibitem{Phi_measure}
		M.~Gazdzicki, S.~Mrowczynski,
		%``A Method to study 'equilibration' in nucleus-nucleus collisions'',
		Z.\ Phys.\ C {\bf 54} (1992) 127.

	\bibitem{KG_MG}
		M.~I.~Gorenstein and K.~Grebieszkow,
		%``Strongly Intensive Measures for Transverse Momentum and Particle Number Fluctuations'',
		Phys.\ Rev.\ C {\bf 89} (2014) 3,  034903
		%[arXiv:1309.7878 [nucl-th]].

\end{thebibliography}
\end{document}